\documentclass{ws-procs9x6}

\begin{document}

\def\changed#1{{\bf #1}}

\title{
SOLITONS IN SUPERSYMMETRIC GAUGE THEORIES: \\
MODULI MATRIX APPROACH
}

\author{MINORU ETO$^\dagger$, YOUICHI ISOZUMI, 
MUNETO NITTA$^{\dagger\dagger}$, KEISUKE OHASHI \\
and NORISUKE SAKAI\footnote{\uppercase{S}peaker at the conference.}}

\address{Department of Physics, Tokyo Institute of 
Technology \\
$^\dagger$Inst. of Physics, 
University of Tokyo \\
$^{\dagger\dagger}$Department of Physics, Hiyoshi, Keio University
}

\begin{abstract}
We review our recent works on solitons in $U(N_{\rm C})$ 
gauge theories with $N_{\rm F}(\geq N_{\rm C})$ 
Higgs fields in the fundamental representation, which possess 
eight supercharges. 
The moduli matrix is proposed as a crucial tool to exhaust 
all BPS solutions, and to characterize all possible 
moduli parameters. 
Since vacua are in the Higgs phase, we find 
domain walls (kinks) and vortices as the only elementary 
solitons. 
Stable monopoles and instantons 
can exist as composite 
solitons with vortices attached. 
Webs of walls are also found as another composite soliton. 
The moduli space of all these elementary as well as 
composite solitons are found in terms of the moduli matrix.
The total moduli space of walls is given by the complex 
Grassmann manifold $
SU(N_{\rm F}) / 
[SU(N_{\rm C})\times SU(N_{\rm F}-{N}_{\rm C}) \times U(1)]$ 
and is decomposed into various topological 
sectors corresponding to boundary conditions 
specified by particular vacua. 
We found charges characterizing composite solitons 
contribute negatively (either positively or negatively) 
in Abelian (non-Abelian) gauge theories. 
Effective Lagrangians are constructed on walls and 
vortices in a compact form. 
The power of the moduli matrix is illustrated 
by an interaction rule of 
monopoles, vortices, and walls, which is difficult to obtain 
in other methods. 
More thorough description of the moduli matrix approach 
can be found in our review article\cite{Eto:2006pg} 
(hep-th/0602170).

\end{abstract}


\bodymatter

\section{Discrete Vacua in Higgs Phase}\label{sec:vacua}

Solitons have been playing a central role in understanding 
nonperturbative effects. 
The solitons are classified by their codimensions.  
Kinks (domain walls), vortices, 
monopoles and instantons are well-known 
typical solitons with codimensions one, two, three and four, 
respectively. 
They carry topological charges classified by certain 
homotopy groups according to their codimensions. 
Moreover, they are also important to construct models of 
the brane-world, where our four-dimensional world is 
realized on a topological defect in higher-dimensional 
spacetime\cite{Horava:1996ma,Arkani-Hamed:1998rs,Randall:1999ee}. 
These topological defects are 
preferably solitons as a solution of field equations. 

When energy of solitons saturates a bound from below, 
which is called the Bogomol'nyi bound, 
they are the most stable among all possible configurations 
with the same boundary condition, 
and automatically satisfy field equations. 
They are called Bogomol'nyi-Prasad-Sommerfield (BPS) 
solitons\cite{Bogomolny:1975de}.  
If a part of supersymmetry (SUSY) is preserved 
in supersymmetric theories, the field 
configuration becomes a BPS state\cite{Witten:1978mh}. 
The representation theory of SUSY shows that 
they are non-perturbatively stable. 
With this fact non-perturbative effects have been 
established in SUSY gauge theories and string theory\cite{Seiberg:1994rs}. 
In supersymmetric theories, BPS solitons often have 
parameters, which are called moduli. 
When they are promoted to fields on the world volume of 
solitons, they become massless fields of the low-energy 
effective theory.


We are primarily interested in $U(N_{\rm C})$ gauge theory 
with $N_{\rm F}$ flavors in the fundamental representation, 
which can be made SUSY theories with eight supercharges. 

Bosonic components of a vector multiplet are a gauge field 
$W_M, M=0, 1, \cdots, d-1$ and a real adjoint scalar field 
$\Sigma_p, p=d, \cdots, 5$ in the adjoint representation. 
Matter fields are represented by hypermultiplets containing 
two $N_{\rm C}\times N_{\rm F}$ matrices 
of complex Higgs (scalar) 
fields $H^1, H^2$ as bosonic components. 
The theory contains a common gauge coupling 
$g$ for $SU(N_{\rm C})$ and $U(1)$ and the Fayet-Iliopoulos 
parameter\cite{Fayet:1974jb} $c$. 
Since one of the hypermultiplet scalar $H^2 =0$ in all of 
our BPS solutions if $c>0$, we ignore it: 
$H^1\equiv H$, $H^2=0$. 
Our (bosonic part of) the Lagrangian is given by 
\begin{eqnarray}
{\cal L} &=& {\cal L}_{\rm kin} - V, 
\label{eq:mdl:total_lagrangian}
\\ 
{\cal L}_{\rm kin} &=& 
{\rm Tr}\left(- {1\over 2g^2}F_{\mu\nu}F^{\mu\nu} 
+\frac{1}{g^2}{\cal D}_\mu\Sigma_p{\cal D}^\mu\Sigma_p 
+{\cal D}^\mu H \left({\cal D}_\mu H\right)^\dagger \right), 
\label{eq:mdl:lagrangian}
\end{eqnarray}
where the covariant derivatives and field strengths are 
defined as 
${\cal D}_\mu \Sigma_p=\partial_\mu\Sigma_p + i[W_\mu, \Sigma_p]$, 
${\cal D}_\mu H=(\partial_\mu + iW_\mu)H$, 
$F_{\mu\nu}=-i[{\cal D}_\mu,\,{\cal D}_\nu]$. 
Our convention for the metric is 
$\eta_{\mu\nu} = {\rm diag}(+,-,\cdots,-)$. 
The scalar potential $V$ is given in terms of 
diagonal mass matrices $M_p$ and a real parameter $c$ as 
\begin{eqnarray}
V&=& 
{\rm Tr}
\Big[
\frac{g^2}{4}
\left(c\mathbf{1}_{N_{\rm C}}
-H  H^{\dagger} 
\right)^2 
+ (\Sigma_p H - H M_p) 
 (\Sigma_p H - H M_p)^\dagger 
\Big] . 
\label{eq:mdl:scalar_pot}
\end{eqnarray}

To obtain domain walls, we need real mass parameters 
$M={\rm diag}(m_1, m_2, \cdots, m_{N_{\rm F}})$. 
Therefore we consider $d=5$ with a single adjoint scalar 
$\Sigma$. 
For simplicity, we choose fully non-degenerate mass: 
$m_A > m_{A+1}$. 
Then the flavor symmetry is broken to 
$U(1)_{\rm F}^{N_{\rm F}-1}$. 
Let us note that a common mass can be absorbed 
into the adjoint scalar $\Sigma$.  
Because of non-degenerate masses, we obtain discrete 
supersymmetric vacua, 
labeled by $N_{\rm C}$ flavors 
$ \langle A_1 A_2 \cdots A_{N_{\rm C}}\rangle$, 
which are called color-flavor locking vacua: 
\begin{eqnarray}
 H^{
rA}=\sqrt{c}\,\delta ^{A_r}{}_A,\quad 
\quad \Sigma ={\rm diag}(m_{A_1}, 
\cdots, 
m_{A_{N_{\rm C}}}). 
\end{eqnarray}
The number of these vacua increases exponentially as 
the number of colors $N_{\rm C}$ and flavors $N_{\rm F}$ 
increases: 
\begin{eqnarray}
{N_{\rm F}! \over (N_{\rm F}-N_{\rm C})!N_{\rm C}!}
\sim e^{N_{\rm F} \log(x^{-x}(1-x)^{-(1-x)})}, \; \; 
x\equiv N_{\rm C}/N_{\rm F} . 
\end{eqnarray}

Since the Higgs $H$ charged under $U(N_{\rm C})$ gauge group 
have nonvanishing values, 
the vacua are in the Higgs phase. 
In the Higgs phase, only walls and vortices are elementary 
solitons, whereas the instantons,  monopoles, and 
(wall-)junctions appear as composite solitons.

\section{$1/2$ BPS Walls}
\begin{figure}[htb]
\begin{center}
\includegraphics[width=7cm,clip]{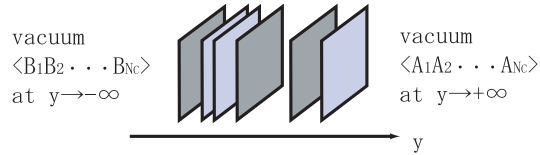}
\end{center}
\caption{
Multi-wall connecting vacua 
$\langle A_1A_2\cdots A_{N_{\rm C}}\rangle$ 
and $\langle B_1B_2\cdots B_{N_{\rm C}}\rangle $. 
}
\label{su2nf}
\end{figure}

To obtain domain wall solutions 
we assume that all fields depend on 
one spatial coordinate, say $y \equiv x^4$  
with $3+1$ dimensional Poincar\'e invariance. 
The {\bf $1/2$ BPS equations for walls}
are obtained 
by requiring the following direction $\varepsilon^i$ 
of SUSY to be preserved\cite{Isozumi:2004jc}: 
$\gamma ^4\varepsilon ^i
=-i(\sigma ^3)^i{}_j\varepsilon ^j$,
\begin{equation}
{\cal D}_y H
=
-\Sigma H
 + H
 M,
\qquad 
\label{eq:wll:bogomolnyi-wall_hyper}
{\cal D}_y \Sigma 
=
{g^2 \over 2}\left(c{\bf 1}_{N_{\rm C}}-H
H
{}^\dagger 
\right) . 
\end{equation}
The topological sector of (multi-)BPS wall configurations is 
labeled by vacua 
$\langle A_1A_2\cdots A_{N_{\rm C}}\rangle$ at $y=\infty$ 
and $\langle B_1B_2\cdots B_{N_{\rm C}}\rangle $ at 
$y=-\infty$ as shown in Fig.~\ref{su2nf}. 

The BPS equations for hypermultiplet 
(the first of Eq.(\ref{eq:wll:bogomolnyi-wall_hyper})) 
can be solved\cite{Isozumi:2004jc} 
by defining an element $S(y)$ of a complexified 
gauge group $GL(N_{\rm C},\mathbf{C})$ as
\begin{equation}
\Sigma + iW_y \equiv S^{-1}(y)\partial_y S(y), 
\label{eq:def_glnc_element} \qquad 
 H
(y) = S^{-1}(y)H_0 e^{My}
. 
\end{equation}
We call the constant $N_{\rm C} \times N_{\rm F}$ matrix 
$H_0$ `` {\bf moduli matrix}''. 

With the above solution, we can rewrite 
the vector multiplet BPS equation into the following 
equation in terms of the gauge invariant quantity 
$\Omega \equiv SS^\dagger$ 
\begin{equation}
 \partial _y\left(\Omega ^{-1}\partial _y\Omega \right)
=g^2c\left(\mathbf{1}_{N_{\rm C}}-\Omega ^{-1}\Omega _0\right),
\qquad \Omega _0\equiv c^{-1}H_0e^{2My}H_0^\dagger. 
\label{eq:master}
\end{equation}
We call this equation ``{\bf master equation}''. 
The index theorem\cite{Sakai:2005sp} 
shows that the number of moduli parameters 
contained in the moduli matrix $H_0$ is just enough, 
implying that the solution of the master equation exists 
and is unique for a given $\Omega_0$. 
The existence and uniqueness have been proved rigorously 
for the case of $U(1)$ gauge theory\cite{Sakai:2005kz}.

Since the solution $S(y)$ of Eq.(\ref{eq:def_glnc_element}) 
has $N_{\rm C}^2$ integration constants, 
two sets $(S, H_0)$ and 
$(S', H_0{}')$ 
give the same $H=S^{-1}H_0e^{My}$, 
if they are related by the following 
global $GL(N_{\rm C},\mathbf{C})$ transformation $V$ 
(called the $V$-transformation), 
\begin{equation}
 S\rightarrow S' = VS,
 \quad 
H_0 \rightarrow H_0{}'=VH_0, 
\qquad 
V\in GL(N_{\rm C},\mathbf{C}). 
\end{equation}
Therefore the genuine moduli parameters of domain walls 
are given by the equivalence class defined by the 
$V$-transformation. 
We thus find that the {\bf total moduli space} for 
(multi-)wall solutions is the complex 
{\bf Grassmann manifold}\cite{Isozumi:2004jc}:
\begin{eqnarray}
 {\cal M}_{N_{\rm F},N_{\rm C}}
&
=
&
\{H_0 | H_0 \sim V H_0, V \in GL(N_{\rm C},\mathbf{C})\} 
\equiv
G_{N_{\rm F},N_{\rm C}}
\nonumber\\
&
\simeq 
& 
{SU(N_{\rm F}) \over 
 SU(N_{\rm C}) \times SU(
N_{\rm F}-N_{\rm C}
) 
 \times U(1)} . 
\label{eq:wll:Gr}
\end{eqnarray} 
This is a compact (closed) 
set of complex dimension 
$N_{\rm C} \tilde N_{\rm C}
\equiv N_{\rm C}(N_{\rm F}-N_{\rm C})$. 
We did not put any boundary conditions at $y \to \pm \infty$ 
to get the moduli space (\ref{eq:wll:Gr}).
Therefore it contains configurations with 
all possible boundary conditions, 
and can be decomposed into 
the sum of topological sectors 
\begin{figure}
\begin{center}
\includegraphics[width=5cm,clip]{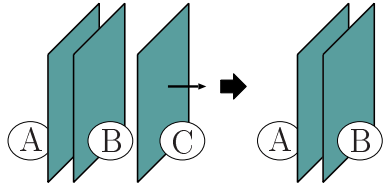}
\caption{ 
A three wall solution connecting vacuum A to C 
through B (left). 
By letting the right-most wall to infinity, we obtain a 
two wall solution connecting vacuum A to B. 
}\label{nit-fig2}
\end{center}
\end{figure}
\begin{eqnarray}
 {\cal M}^{\rm total}_{\rm wall} 
 = \sum_{\rm BPS} {\cal M}^{\langle A_1,\cdots, A_{N_{\rm C}} \rangle 
        \leftarrow \langle B_1,\cdots, B_{N_{\rm C}} \rangle} . 
  \label{eq:wll:decom}
\end{eqnarray}
As shown in Fig.~\ref{nit-fig2}, by sending one of the wall 
to infinity, we obtain one less walls. 
Namely the boundaries of a topological sector consists of 
topological sectors with one less wall. 
It is interesting to observe that this natural compactification 
of the moduli space of walls leads to the compact total 
moduli space (\ref{eq:wll:Gr}) for the 1/2 BPS solutions, 
if we add vacua as points to compactify the manifold.

Components of the moduli matrix $H_0$ represent weights of 
the vacua. 
\begin{figure}
\begin{center}
\includegraphics[width=8cm,height=2cm]
{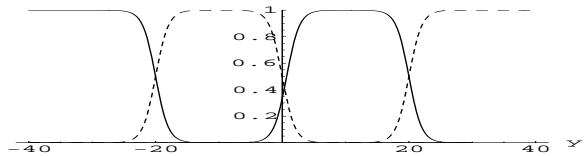}
\caption{ Rapid change of hypermultiplets indicates the 
positions of walls. \label{fig:wall_position}}
\end{center}
\end{figure}
For instance, the moduli matrix for the $U(1)$ gauge 
theory can be parametrized by 
$H_0=(e^{r_1}, e^{r_2}, \cdots, e^{r_{N_F}})$. 
Then the hypermultiplets are given by 
\begin{eqnarray}
H=S^{-1}H_0 e^{My}=S^{-1} (e^{r_1+m_1y}, \cdots, e^{r_{N_F}+m_{N_F}y})
  \label{eq:wll:u1}
\end{eqnarray}
We see that wall separating $i$- and $i+1$-th vacua 
is located where the magnitudes of the $i$- and $i+1$-th 
components become equal as illustrated in 
Fig.~\ref{fig:wall_position}. 
The wall position $y$ is 
\begin{eqnarray}
{\rm Re}r_{i}+m_{i}y \sim {\rm Re}r_{i+1}+m_{i+1}y 
\to 
y=-{\rm Re}(r_{i}-r_{i+1})/(m_i-m_{i+1}) 
. 
\end{eqnarray}
The imaginary part ${\rm Im}(r_{i}-r_{i+1})$ gives the 
relative phase of the two adjacent vacua. 
We see that there are $N_{\rm F}-1$ walls maximally. 
Similarly, the number of walls in non-Abelian $U(N_{\rm C})$ 
gauge theory is given by $N_{\rm C}(N_{\rm F}-N_{\rm C})$, and 
each wall carries two moduli, position and 
relative phase of adjacent vacua.

The low-energy effective Lagrangian on domain walls 
is given by promoting the moduli parameters in the moduli 
matrix $H_0$ to 
fields on the world volume of the soliton and by assuming 
the weak dependence on the world volume coordinates. 
We assume the slow-movement of moduli fields compared 
to the two typical mass scales $g\sqrt{c}$ 
and $\Delta m$ of the wall of hypermultiplets 
\begin{eqnarray}
 \lambda \ll {\rm min}(\Delta m, g\sqrt{c} ).
\label{eq:efa:slow_move}
\end{eqnarray}
We find it extremely useful to use the superfield formalism 
maintaining the preserved four SUSY manifest. 
The {\bf effective Lagrangian} for the 
$1/2$ BPS domain walls is given in terms of 
the solution $\Omega _{\rm sol}(y, \phi, \phi^*)$ of 
the master equation
\begin{eqnarray}
{\cal L}
=
-T_{\rm w}+\int d ^4\theta K(\phi ,\phi ^*)
+{\rm higher}\; {\rm derivatives}, 
\end{eqnarray}
where $T_{\rm w}$ is the tension of the domain wall, 
and $K$ is the {\bf K\"ahler potential} of moduli fields 
$\phi, \phi^*$, given by\cite{Eto:2006uw} 
\begin{eqnarray}
\!\!\!
K(\phi ,\phi ^*)
\! =\!\!
\int \! dy \left[
c\log{\rm det}\Omega 
+c{\rm Tr}\left(\Omega _0\Omega ^{-1}\right)
+{1\over 2g^2}{\rm Tr}
\left({\Omega }^{-1}\partial _y\Omega \right)^2 
\right]
\Big|_{\Omega =\Omega _{\rm sol}
}
\end{eqnarray}
One of the merit of our superfield formulation is that 
K\"ahler potential can be obtained directly without 
going through K\"ahler metric and integrating it. 
We find that the K\"ahler potential serves as the 
action for $\Omega$ to obtain the master equation 
(\ref{eq:master}). 

Let us make some comments.
If we take the strong gauge coupling limit 
$g^2c/(\Delta m)^2 \gg 1$, the model 
becomes a nonlinear sigma model\cite{Arai:2002xa} 
and the 
master equation (\ref{eq:master}) can be solved 
 algebraically 
\begin{eqnarray}
 \Omega
 = \Omega_0 \equiv 
 c^{-1}H_0 e^{2My}H_0^\dagger. 
\end{eqnarray} 
The domain wall configuration in our system 
can be realized as a bound state of 
kinky D$p$-brane and D($p+4$)-branes in 
the type II string theories\cite{Eto:2004vy}. 
By doing so ample dynamics of walls 
have been uncovered. 
We have found that the moduli space of domain walls is 
generally the Lagrangian submanifold of the vacuum manifold of 
corresponding massless model\cite{Eto:2005wf}.

\section{$1/2$ BPS Vortices
} 

Vortices can exist in $5+1$ dimensions or lower. 
In particular they carry 
non-Abelian orientational moduli in 
massless theory\cite{Hanany:2003hp,Auzzi:2003fs}  
as instantons.
For simplicity let us consider the case of 
$N_{\rm F}=N_{\rm C}=N$. 
Taking the Lagrangian (\ref{eq:mdl:lagrangian}) 
in $5+1$ dimensions, 
and requiring the half of SUSY to be preserved, 
we obtain 
the $1/2$ BPS equations for vortices as 
\begin{eqnarray}
 0 = {\cal D}_1 H + i {\cal D}_2 H, \quad 
 0 = F_{12} + {g^2 \over 2} (c {\bf 1}_N - H H^\dagger). 
\end{eqnarray}
Hypermultiplet BPS equation can be easily solved in terms 
of a complexified gauge transformation 
$S(z, \bar z) \in GL(N_{\rm C},\mathbf{C})$ 
and the holomorphic moduli matrix $H_0(z)$\cite{Isozumi:2004vg,Eto:2005yh}
\begin{eqnarray}
 H =S^{-1} H_0(z),
  \quad
 W_1+iW_2 =-i2S^{-1}{\bar \partial}_z S, 
\quad 
z \equiv x^1+ix^2. 
\end{eqnarray}
The vector multiplet BPS equation can be transformed to the 
following master equation 
\begin{eqnarray}
 \partial_z (\Omega^{-1} \bar \partial_z \Omega ) 
 = {g^2 \over 4} (c{\bf 1}_N - \Omega^{-1} H_0 H_0^\dagger). 
\end{eqnarray}
The solutions of the BPS equation saturate the 
BPS bound for the energy density for 
{\bf vorticity} $k \in {\bf Z}_{\geq 0}$ 
\begin{eqnarray}
 T \equiv -c \int d^2 x\ {\rm Tr} F_{12} = 2 \pi c k 
 = -i {c\over 2} \oint dz\ \partial {\rm log}({\rm det}H_0) 
 + {\rm c.c.}, 
\end{eqnarray}
with the boundary condition 
$
{\rm det}(H_0) \sim z^k$ at $z \to \infty $. 
The moduli matrices $H_0(z)$ related by the 
$V$-transformation give identical physical fields : 
$ H_0 \to V H_0$, $S \to V S$, $V = V(z) \in GL(N,{\bf C}), 
\det V =$ const.$\neq 0$. 
Therefore, the moduli space for vortices is found as
\begin{eqnarray}
{\cal M}_{k,N}={\left\{H_0(z)|H_0(z)\in M_N, 
{\rm deg}\, {\rm det}(H_0(z)) = k \right\}\over 
\left\{V(z)|V(z)\in M_N, 
{\rm det}V(z)={\rm const.}\not=0 \right\}} . 
\end{eqnarray}

The generic points of moduli space has 
${\rm dim}({\cal M}_{N,k})=2k N$ 
and can be represented by 
\begin{eqnarray}
H_0 = 
\left(
\begin{array}{cc}
{\bf 1}_{N-1} & - \vec R(z) \\
0 & P(z)
\end{array}
\right), \quad P(z) = \prod_{i=1}^k(z-z_i). 
  \label{eq:modmatvor}
\end{eqnarray}
Moduli space of a single vortex $k=1$ 
is given\cite{Hanany:2003hp,Auzzi:2003fs} by 
${\cal M}_{N,k=1}$ $\simeq$ 
${\bf C} \times {\bf C}P^{N-1}$ 
and is represented by the moduli matrix (\ref{eq:modmatvor})
with $\vec {R}^T = (b_1 ,\cdots, b_{N-1})$ and $P(z) = z-z_0$.
Moduli space of $k$ separated vortices is given by 
a symmetric product 
$\left({\bf C}\times{\bf C}P^{N-1}\right)^k / \mathfrak{S}_k$. 
The orbifold singularities of this are 
appropriately resolved in the full moduli space\cite{Eto:2005yh}. 
We also find that 
the K\"ahler quotient construction\cite{Hanany:2003hp} 
can also be obtained from our moduli matrix 
by a change of basis. 
Superfield formulation with the slow-movement expansion 
readily yields the effective Lagrangian on the world volume 
of vortices\cite{Eto:2006uw}. 
The duality between vortices and walls 
has been discussed\cite{Eto:2006mz}.

\section{1/4 BPS Webs of Domain Walls} 

The direction of BPS walls are related to the phase of 
the hypermultiplet masses. 
If we have {\bf complex masses} $\mu_A=m_A+in_A$, 
we can obtain two or more non-parallel walls, which can lead to 
{\bf wall junctions}\cite{Gibbons:1999np}. 
We consider the Lagrangian (\ref{eq:mdl:lagrangian}) 
in $3+1$ dimensions, 
since complex masses can be realized in $3+1$ dimensions 
or lower 
\begin{eqnarray}
M_1={\rm diag}\left(m_1,m_2,\cdots,m_{N_{\rm F}}\right), 
\quad 
M_2 = {\rm diag}\left(n_1,n_2,\cdots,n_{N_{\rm F}}\right). 
\end{eqnarray}
The wall junctions can be realized as a solution of 
the following $1/4$ BPS equations\cite{Eto:2005cp} 
\begin{eqnarray}
F_{12} = i \left[\Sigma_1,\Sigma_2\right],\quad
{\cal D}_1\Sigma_2 = {\cal D}_2\Sigma_1,\quad
{\cal D}_{\alpha} H = HM_{\alpha} - \Sigma_{\alpha} H , 
\label{eq:web_bps1}
\end{eqnarray}
\begin{eqnarray}
{\cal D}_1\Sigma_1 + {\cal D}_2\Sigma_2 = 
{g^2 \over 2}(c {\bf 1}_{N_{\rm C}} -H H^\dagger)
.
\label{eq:web_bps2}
\end{eqnarray}
The solutions saturate the Bogomol'nyi bound for the energy 
density 
\begin{eqnarray}
{\cal E} 
&\ge& {\cal Y}+ {\cal Z}_1 + {\cal Z}_2 
+ \sum_{\alpha=1, 2}\partial_\alpha J_\alpha, 
\quad 
J_\alpha \equiv {\rm Tr }\Big[ 
H(M_\alpha H^\dagger - H^\dagger \Sigma_\alpha)\Big] , 
\end{eqnarray}
\begin{eqnarray}
{\cal Y} \equiv \frac{2}{g^2}
\partial_\alpha{\rm Tr}\left(\epsilon^{\alpha\beta}
\Sigma_2{\cal D}_\beta\Sigma_1\right),\quad
{\cal Z}_1 \equiv c \partial_1 {\rm Tr} \Sigma_1,\quad
{\cal Z}_2 \equiv c \partial_2 {\rm Tr} \Sigma_2 . 
\end{eqnarray}

The first two equations in Eq.(\ref{eq:web_bps1}) 
assures the integrability 
of the last one in Eq.(\ref{eq:web_bps1}), which 
is solved by\cite{Eto:2005cp} 
\begin{eqnarray}
H = S^{-1}H_0e^{M_1x^1 + M_2x^2},\quad
W_{\alpha} - i\Sigma_{\alpha} 
= -iS^{-1}\partial_{\alpha} S,\quad \alpha=1,2 . 
\end{eqnarray}
The remaining BPS equation (\ref{eq:web_bps2}) can be rewritten 
in terms of the gauge invariant quantity 
$\Omega\equiv SS^\dagger$ as the master equation
\begin{eqnarray}
\sum_{\alpha=1,2} 
 \partial_{\alpha} \left(\partial_{\alpha}\Omega\Omega^{-1}\right)
= cg^2\left({\bf 1}_{N_{\rm C}} - 
c^{-1}H_0e^{2(M_1x^1+M_2x^2)}H_0^\dagger 
\Omega^{-1}\right) , 
\end{eqnarray}

\begin{figure}[ht]
\begin{center}
\begin{tabular}{ccc}
\includegraphics[height=2.5cm]{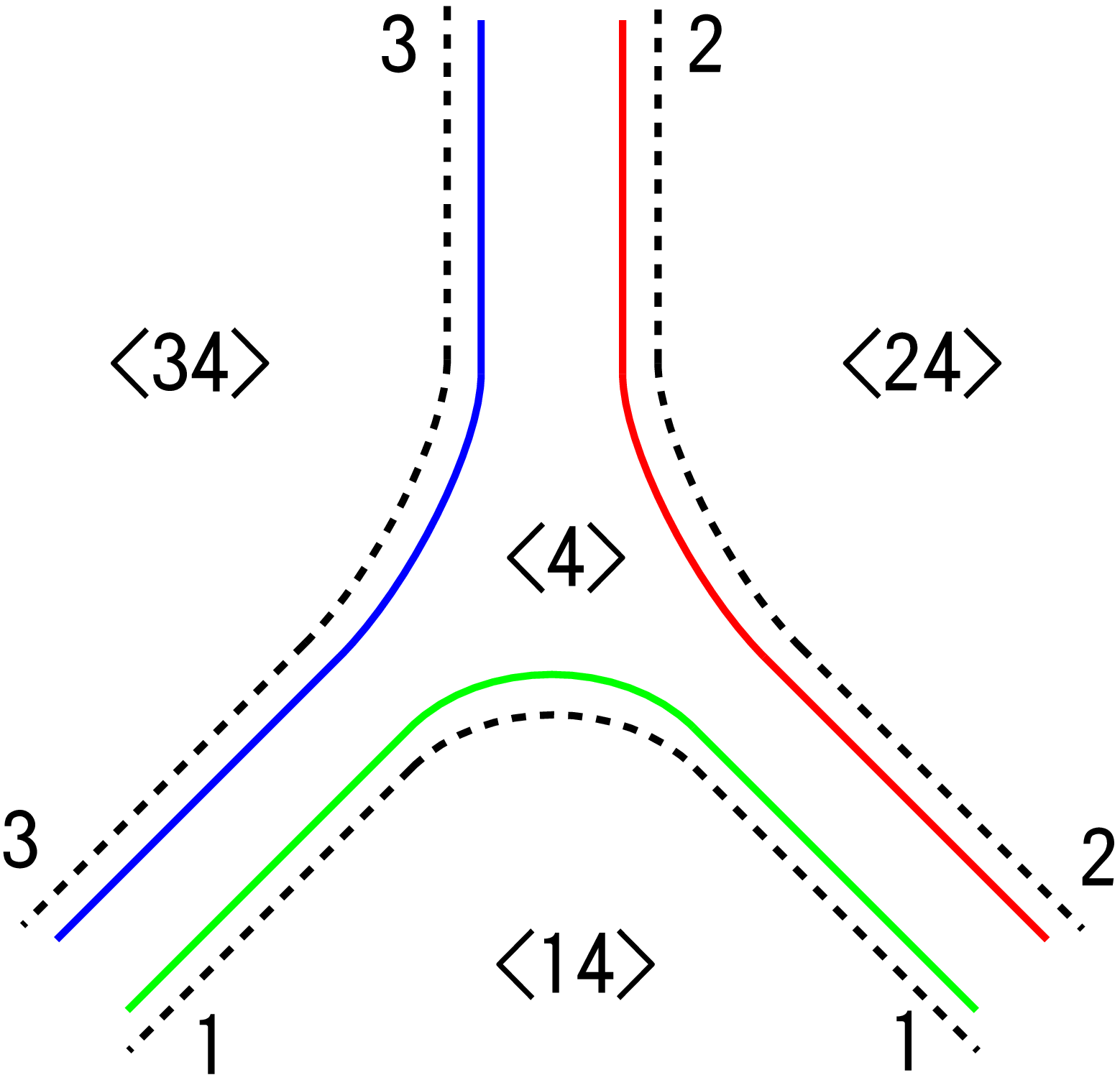}
&\qquad\qquad&
\includegraphics[height=2.5cm]{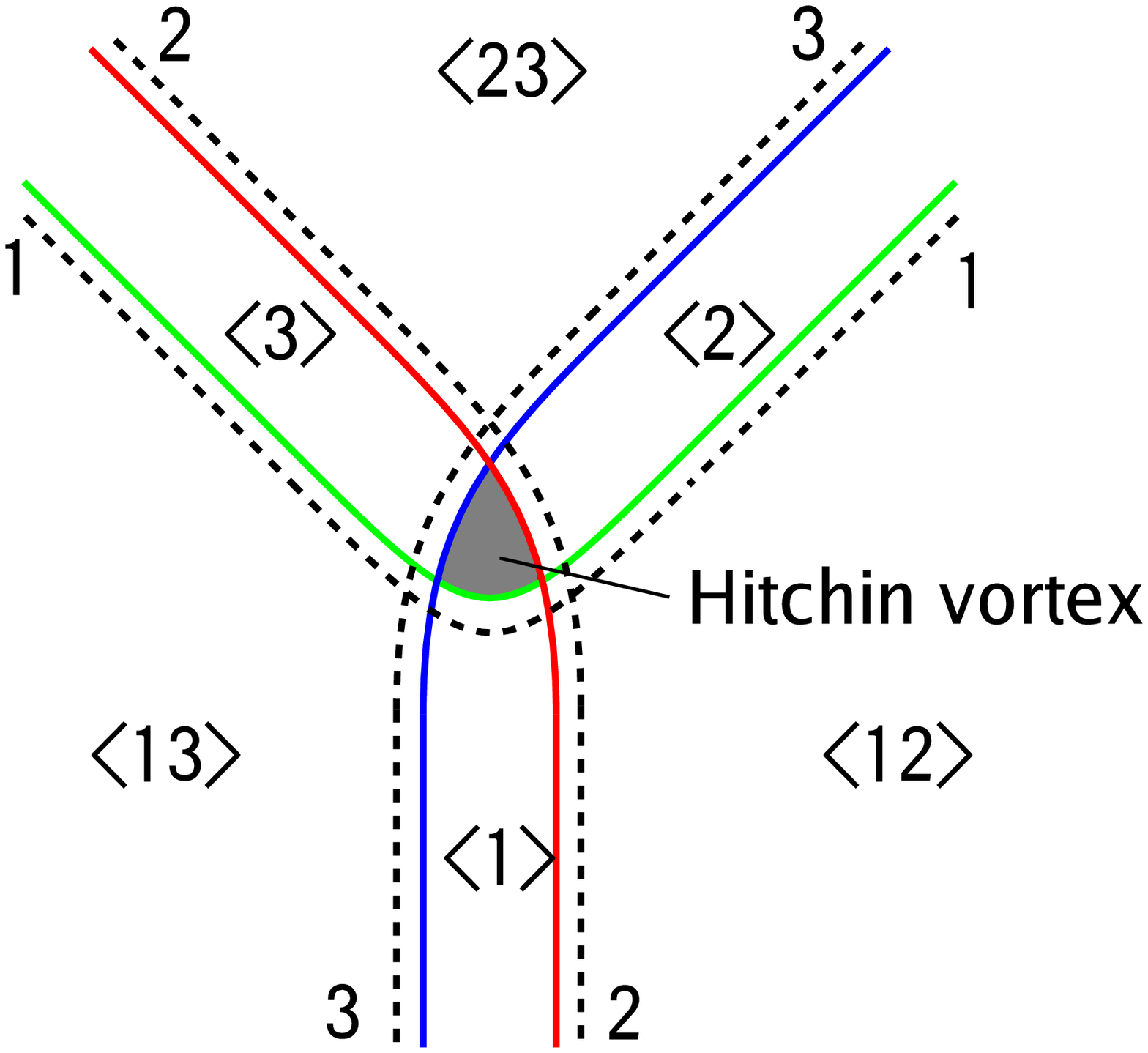}\\
\footnotesize{ 
(a) Abelian junction} 
& & \footnotesize{
 (b) non-Abelian junction}
\vspace*{-.3cm}
\end{tabular}
\caption{ 
Internal structures of the junctions
 with $g \sqrt c \ll |\Delta m+i\Delta n|$}
\label{hitchin}
\vspace*{-.3cm}
\end{center}
\end{figure}

The total moduli space of $1/4$ BPS equations 
(\ref{eq:web_bps1}), (\ref{eq:web_bps2}) 
can be decomposed into $1/4$, $1/2$, and $1/1$ BPS sectors 
\begin{eqnarray}
{\cal M}^{\rm webs}_{\rm tot} 
&\simeq & G_{N_{\rm F},N_{\rm C}} =
\{H_0\ |\ H_0\sim VH_0,\ V\in GL(N_{\rm C},{\bf C})\}
\nonumber \\
&=& {\cal M}^{\text{webs}}_{1/4}
\ \bigcup\ 
{\cal M}^{\text{walls}}_{1/2}
\ \bigcup\ 
{\cal M}^{\text{vacua}}_{1/1}. 
\end{eqnarray}
We find that Abelian gauge theory gives only Abelian junctions 
with the negative junction charge ${\cal Y}<0$, 
whereas non-Abelian gauge theory 
gives non-Abelian junction with positive junction charge ${\cal Y}>0$ 
in addition to Abelian junction.
Physical interpretation of the positive junction charge is 
the presence of the Hitchin vortex residing at the junction 
as illustrated in Fig.~\ref{hitchin}. 
We also find that there are cases with vanishing junction 
charge ${\cal Y}=0$ corresponding to the intersections of penetrable 
walls.

\begin{figure}[ht]
\begin{center}
\begin{tabular}{ccc}
\includegraphics[height=2.5cm]{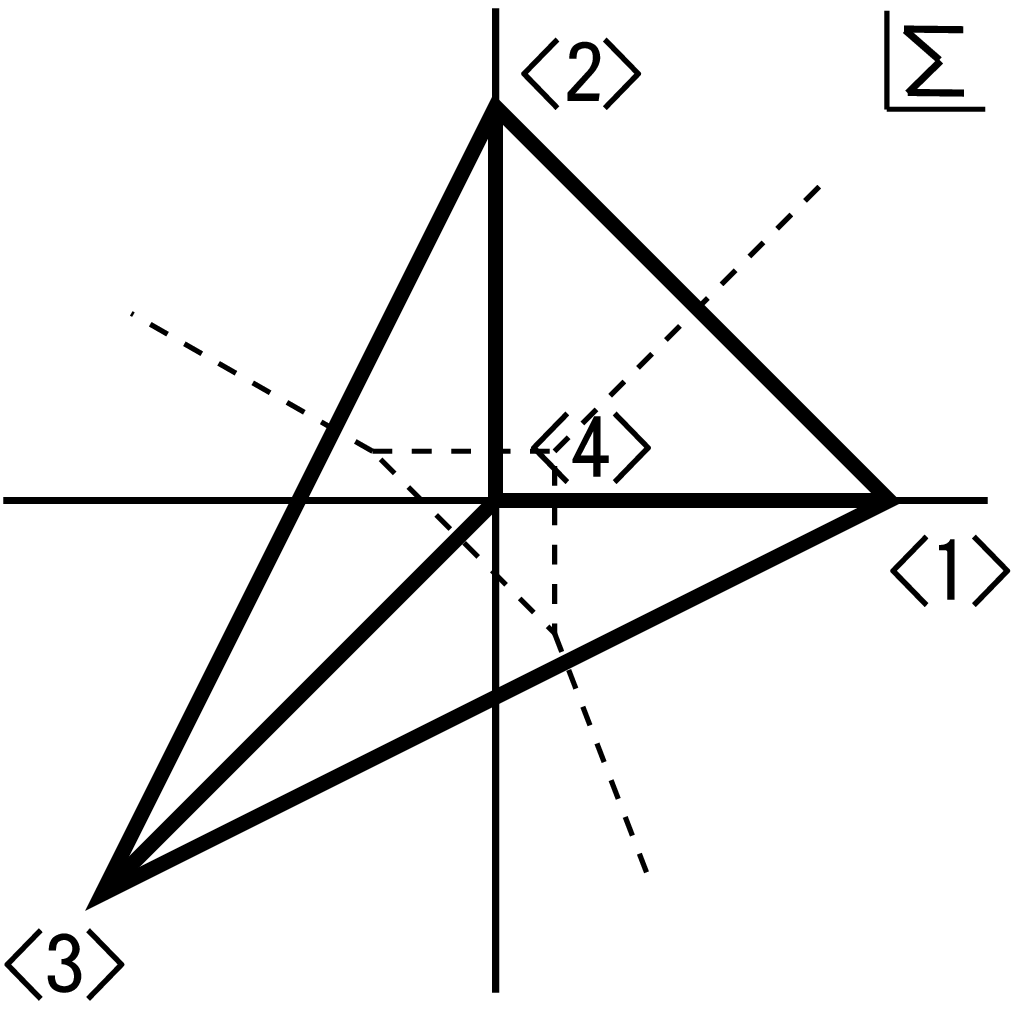} & \qquad\qquad&
\includegraphics[height=2.5cm]{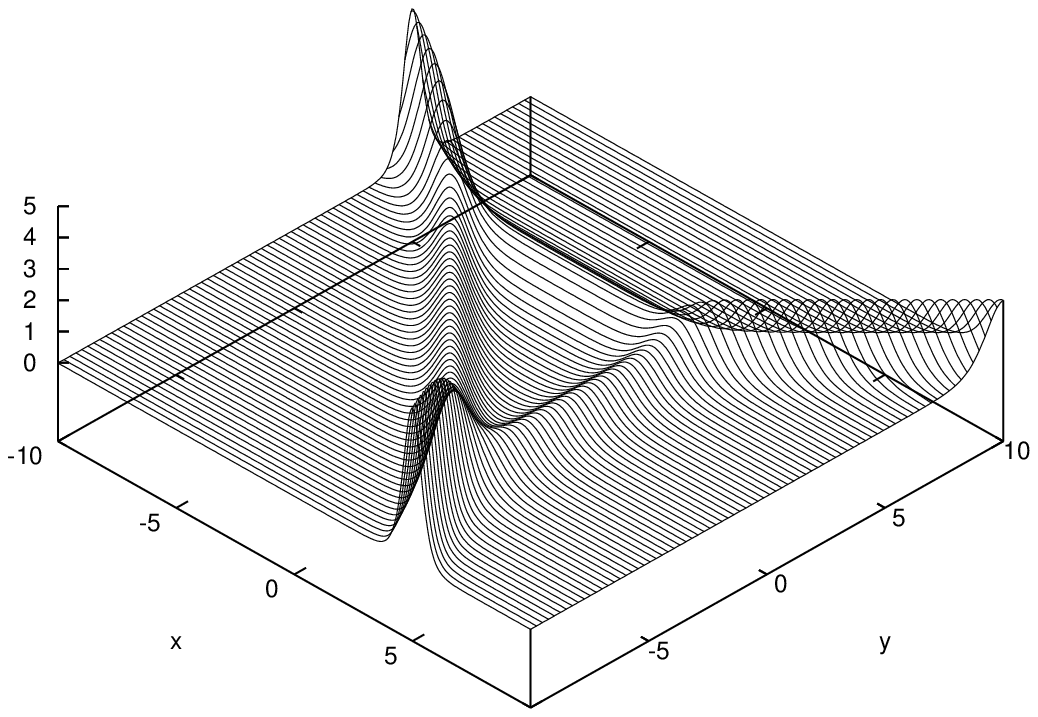}\\
{\footnotesize  
(a) grid diagram} 
&&{\footnotesize 
(b) energy density ($g^2\to\infty$)}
\end{tabular}
\caption{ 
Abelian junction with $1$ loop and 
 $3$ external walls in  $N_{\rm C}=1, N_{\rm F}=4$ model.
}
\label{cp3_loop}
\vspace*{-0.8cm}
\end{center}
\end{figure}

We find that the {\bf normalizable moduli} of web of walls 
are given by {\bf loops} in web as shown in 
Fig.~\ref{cp3_loop}. 
The moduli matrix of 
$N_{\rm F}=4,N_{\rm C}=1$ case with 
$M = {\rm diag}(1,i,-1-i,0)$ can be 
parametrized by 
\begin{eqnarray}
H_0 = \sqrt c \left( e^{a_1+ib_1},e^{a_2+ib_2},e^{a_3+ib_3},
e^{a_4+ib_4}\right), 
\end{eqnarray}
with $a_j+ib_j, j=1,2,3$ as external wall moduli, 
and $a_4+ib_4$ as the loop moduli, corresponding to 
the normalizable mode. 
Grid diagrams are found to be useful to specify the 
moduli of the web of walls as illustrated in Fig.~\ref{cp3_loop}. 
A brane configuration is proposed in \cite{Eto:2005mx}.

\section{
1/4 BPS Monopoles (Instantons) inside a vortex
} 

\begin{figure}[htb]
\begin{center}
\includegraphics[width=4.5cm,clip]{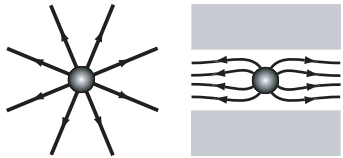}
\end{center}
\caption{ Monopole in Coulomb phase (left) and in Higgs phase (right). }
\label{fig:monopole}\end{figure}

Vacua outside of monopoles are in the Coulomb phase 
with unbroken $U(1)$ gauge group. 
In our $U(N_{\rm C})$ gauge theory with 
$N_{\rm F} (\ge N_{\rm C})$ flavors of hypermultiplets in 
the fundamental representation, vacua are in the Higgs phase. 
If we place a monopole in the Higgs phase, 
magnetic flux emanating from the monopole is squeezed 
into vortices, as illustrated in Fig.~\ref{fig:monopole}. 
Therefore {\bf monopoles in the Higgs phase} 
become composite of monopoles and vortices\cite{Tong:2003pz}. 
Monopoles depend on $x^1, x^2, x^3$ coordinates 
and preserve $1/2$ SUSY defined by 
$\gamma ^{123}\varepsilon ^i=\varepsilon ^i$. 
Vortices along $x^3$-axis preserve another $1/2$ SUSY 
defined by 
$\gamma ^{12}(i\sigma _3)^i{}_j\varepsilon ^j=\varepsilon ^i$. 
If they coexist as a monopole in the Higgs phase, 
$1/4$ of SUSY is preserved. 
$\gamma ^3(i\sigma _3)^i{}_j\varepsilon ^j
=\varepsilon ^i$. 
We found that this $1/4$ SUSY precisely allows 
domain walls 
perpendicular to the vortices\cite{Isozumi:2004vg}. 

Let us consider $U(N_{\rm C})$ gauge theory in $4+1$ 
dimensions and assume field configurations of 
monopole-vortex-wall composite to depend on 
$x^m\equiv (x^1, x^2, x^3)$ and 
Poincar\'e invariance in $x^0,x^4$ space. 
With the above preserved $1/4$ SUSY, we obtain the 
$1/4$ BPS equations : 
\begin{eqnarray}
{\cal D}_3 \Sigma
&=&{g^2 \over 2}
\left(c{\bf 1}_{N_{\rm C}}-HH{}^\dagger\right)
 {+F_{12}} ,
\quad {\cal D}_3 H 
=
-\Sigma H +H M,
\end{eqnarray}
which amounts to the contribution of vortex magnetic 
field $F_{12}$ added to the wall BPS equation. 
These are supplemented by the BPS equations for vortices 
\begin{eqnarray}
0&=&{\cal D}_1 H +i{\cal D}_2 H,
\quad 
0
=
F_{23}
-{\cal D}_1 \Sigma,\quad 
0=F_{31}
-{\cal D}_2  
\Sigma . \label{eq:wvmBPS2}
\end{eqnarray}
The solutions of these BPS equations saturate the 
{\bf BPS bound} of the energy density 
\begin{eqnarray}
{\cal E}\geq t_{\rm w}+t_{\rm v}+t_{\rm m}+\partial _mJ_m , 
\end{eqnarray}
where $J_m$ is the current that does not contribute to the 
topological charge, and 
$t_{\rm w},\,t_{\rm v}$ and $t_{\rm m}$ are 
energy densities for {\bf walls}, {\bf vortices} and 
{\bf monopoles} 
\begin{eqnarray}
  {t_{\rm w}}= {c\partial _3 {\rm Tr}(\Sigma)},\quad 
  {t_{\rm v}}= {-c{\rm Tr} ( F_{12})},\quad 
  {t_{\rm m}}=
 {\frac{2}{g^2}\partial _m {\rm Tr}( \frac12 \epsilon^{mnl} \! 
F_{nl}\Sigma )}. 
\end{eqnarray}

Integrability condition 
$[{\cal D}_1+i{\cal D}_2,\,{\cal D}_3+\Sigma]=0$ 
coming from the second and third equations in (\ref{eq:wvmBPS2})
assures the existence of an invertible complex matrix 
function $S(x^m)\in GL(N_{\rm C},\mathbf{C})$ 
defined by\cite{Isozumi:2004vg} 
\begin{eqnarray}
({\cal D}_3+\Sigma)S^{-1}=0 \rightarrow 
\Sigma + iW_3 \equiv S^{-1}\partial_3 S, 
\end{eqnarray}
\begin{eqnarray}
({\cal D}_1+i{\cal D}_2)S^{-1}=0 \rightarrow 
W_1+iW_2\equiv -2iS^{-1}\bar \partial S , 
\end{eqnarray}
where $z \equiv x^1 + i x^2$, and 
$\bar \partial\equiv \partial/\partial z^*$. 
With this matrix function, the 
BPS Eq.~for hypermultiplet is solved by 
\begin{eqnarray}
 H = S^{-1}(z,z^*, x^3)H_0(z) e^{M x^3}, 
\label{sol-H}
\end{eqnarray}
where the {\bf moduli matrix} 
$H_0(z)$: $N_{\rm C} \times N_{\rm F}$ 
matrix is a holomorphic function of $z$. 
\begin{figure}[htb]
\begin{center}
\includegraphics[width=6.0cm
]{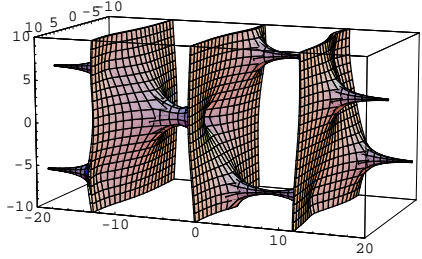}
%
%
\caption{\label{fig:wormhole} 
Surfaces defined by the same energy density 
 with 
$t_{\rm w}+t_{\rm v}=0.5c$. 
}
\end{center}
\end{figure}
The remaining BPS equation is rewritten into a 
master equation for $\Omega \equiv SS^\dagger $ 
with $\Omega_0\equiv H_0 \,e^{2My} H_0{}^\dagger$ 
as input data\cite{Isozumi:2004vg} 
\begin{eqnarray}
4\partial(\Omega ^{-1} \bar \partial \Omega) 
+\partial _3(\Omega ^{-1} \partial _3\Omega )
= g^2 
\left(c - \Omega ^{-1}\Omega_0
\right) . 
\end{eqnarray}

We can obtain exact solutions at strong coupling limit: 
$g^2\rightarrow\infty$, since the 
master equation reduces to an algebraic equation 
\begin{eqnarray}
 \Omega
= \Omega_{0} 
\equiv c^{-1}H_0 e^{2My}H_0^\dagger . 
\end{eqnarray}
Our construction produces rich contents, even for 
$U(1)$ gauge theory whose moduli matrix is given by 
\begin{eqnarray}
H_0(z)=\sqrt{c}\left(f^1(z),\,\dots,f^{N_{\rm F}}(z)\right), 
\quad 
\Omega =\sum_{A=1}^{N_{\rm F}}|f^A(z)|^2e^{2m_Ax^3} . 
\end{eqnarray}
Nonconstant $f^A(z)$ can be interpreted as wall 
positions to depend on $z$. 
In particular, 
walls are bent to form {\bf vortices}, if 
$f^A(z)$ has {\bf zeroes}. 
If $f^A(z) \propto (z-z^A_{\alpha })^{k^A_{\alpha }}$, 
we obtain vorticity $k_{\alpha}^A$ 
at $z=z^A_{\alpha}$ on the $A$-th wall. 
An illustrative configuration of monopoles-vortices-walls 
is given in Fig.~\ref{fig:wormhole}. 
A {\bf monopole in Higgs phase} is realized as a 
{\bf kink on a vortex}, whereas 
{\bf instantons inside a vortex} is realized as a 
{\bf vortex on a vortex}\cite{Eto:2004rz}.

The moduli matrix approach is powerful enough to establish 
interaction rules of monopoles, vortices, and walls. 
In $U(2)$ gauge theory with $N_{\rm F}=3$ flavors, 
we can list up all possible moduli matrices for 
a single vortex in both sides of a wall 
($M = {\rm diag}(m_1,m_2,m_3)$ ordered as 
$m_1>m_2>m_3$) 
\begin{eqnarray}
\left(
\begin{array}{ccc}
z-z_2 & a_2(z-z_3) & 0\\
0 & 0 & 1
\end{array}
\right)e^{Mx^3},
\quad 
\left(
\begin{array}{ccc}
1 & a_3 & 0 \\
0 & 0 & z-z_4
\end{array}
\right)e^{Mx^3},
\label{eq:moduli_matrix1}
 \\
\left(
\begin{array}{ccc}
1 & a_1 & b \\
0 & 0 & z-z_1
\end{array}
\right)e^{Mx^3} \sim
\left(
\begin{array}{ccc}
\frac{1}{a_1}(z-z_1) & z-z_1 & 0 \\
\frac{1}{b} & \frac{a_1}{b} & 1
\end{array}
\right)e^{Mx^3}. 
\label{eq:moduli_matrix2}
\end{eqnarray}
\begin{figure}[ht]
\begin{center}
\includegraphics[height=3cm]{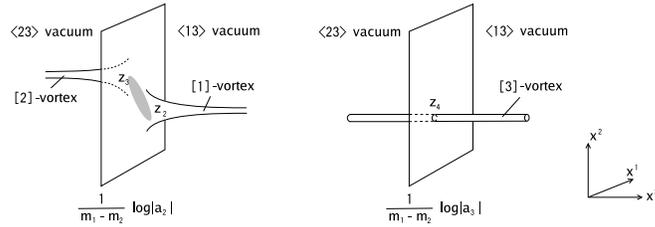}
\caption{
The vortex positions can separate 
on domain wall}
\label{fig:wvm:wvm_bm23}
\vspace*{-.3cm}
\end{center}
\end{figure}
Moduli matrices in (\ref{eq:moduli_matrix1}) are essentially 
those in $U(1)$ gauge theory. 
The first one has two vortices at $z=z_2, z_3$ stretching to 
opposite directions as in the left of Fig.~\ref{fig:wvm:wvm_bm23}. 
The second one represents a single vortex penetrating the 
wall as in the right of Fig.~\ref{fig:wvm:wvm_bm23}. 
We see that vortices can end on a wall in different positions 
as long as monopole does not sit on any of the vortices. 
The moduli matrix in Eq.(\ref{eq:moduli_matrix2}) 
is intrinsically non-Abelian and gives the configuration 
depicted in Fig.~\ref{fig:wvm:wvm_bm1}. 
We find that monopole can penetrate through a wall as long as 
the position of vortices on both sides of the wall coincide. 
Vortices on a wall can separate only when the monopole is 
removed to infinity. 
\begin{figure}[ht]
\begin{center}
\includegraphics[height=3cm]{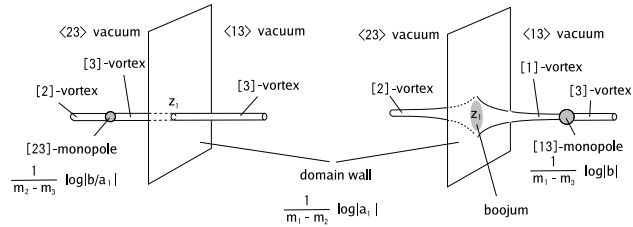}
\caption{
Monopole can go through wall only if 
vortex positions on the wall coincide. 
}
\label{fig:wvm:wvm_bm1}
\vspace*{-.3cm}
\end{center}
\end{figure}

We find that these $1/4$ BPS composite solitons are 
related by the Scherk-Schwarz dimensional reduction 
from $5+1$ dimensions 
to $4+1$ or $3+1$ dimensions as the table below.  
Dyonic extension of these solitons are also discussed \cite{Lee:2005sv}.

\vspace{0.3cm}

\begin{tabular}{c|c|c}
\hline
 dim $\setminus$ charge & positive & negative \\ \hline 
$d=5,6$ instanton & Instanton inside vortex & 
   Intersecton\cite{Eto:2004rz} \\
$d=4,5$ monopole   & Monopole attached by vortices & 
   Boojum\cite{Isozumi:2004vg,Sakai:2005sp}\\
$d=3,4$ Hitchin    & Non-Abelian wall junction & Abelian wall 
   junction\cite{Eto:2005cp}\\
\hline
\end{tabular}

\section{Conclusion}
\begin{enumerate}
\item
The BPS solitons are constructed in 
SUSY $U(N_{\rm C})$ gauge theories 
with $N_{\rm F}$ hypermultiplets in the fundamental representation. 
\item
Total moduli space of the non-Abelian walls is given 
by a compact complex Grassmann manifold 
described by the moduli matrix $H_0$. 
\item
A general formula for the effective Lagrangian is obtained. 
\item
Webs of domain walls are obtained. 
There are Abelian and non-Abelian junctions 
of walls in non-Abelian gauge theory. 
Normalizable moduli of the web of walls are associated 
with loops of walls. 
\item
Composite $1/4$ BPS solitons in Higgs phase are 
systematically obtained by Scherk-Schwarz dimensional 
reduction: instanton-vortex-vortex, 
wall-vortex-monopole, webs of walls.
\item
All possible $1/4$ BPS solutions are obtained exactly 
and explicitly in the  {strong gauge coupling limit}. 
\end{enumerate}

\section*{Acknowledgements}
We would like to thank Toshiaki Fujimori, 
Kazutoshi Ohta, Yuji Tachikawa, David Tong, and Yisong Yang 
for collaborations in various stages. 
This work is supported in part by Grant-in-Aid for 
Scientific Research from the Ministry of Education, 
Culture, Sports, Science and Technology, Japan No.17540237 
(N.~S.). 
The work of K.~O.~(M.~E.~and Y.~I.) is 
supported by Japan Society for the Promotion 
of Science under the Post-doctoral (Pre-doctoral) Research 
Program.

\bibliographystyle{ws-procs9x6}


\end{document}